# A new answer to the Needham Question,
# or Who, how and why did invent the modern physics?

*Gennady Gorelik*

Center for Philosophy and History of Science, Boston University, USA

gorelik@bu.edu

*Abstract* - The cultural infrastructure that let Galileo invent the modern physics is discussed. The key new element of modern physics was firm belief in its fundamental structure, which could be expressed in the double postulate: 1) There are fundamental axioms that all the physical laws could be deduced from; those axioms are not evident, as invisible as the underground foundation stones, or, in Latin, *fundamentum*; 2) The human mind is able to probe into this fundamental level of the Universe to understand its working, and any person is free to contribute in the process of this probing and understanding. Experimentalism and mathematization were just the tools to realize this belief. The modern science was invented in the time when the Bible played the most prominent cultural role in its history due to Gutenberg and Reformation. All the originators of the modern physics were profound biblical believers, and for them the fundamental double postulate was supported by the basic postulates of Biblical worldview.

## The Needham Question

Science, as the process of gaining knowledge about nature, has no certain date and place of birth. For millennia it was fused with practical technology, and the two words could be fused into "technowledge".

However as far as physics is concerned there is consensus that something very important happened in the 17$^{th}$ century that deserves to be called the birth of modern physics. Sure, Archimedes (III BC) was a physicist and such a good one, that Galileo named him "*the most divine*

*Archimedes*". But it was Galileo who had invented something profoundly new to make Einstein to title him "*the father of modern physics*".[1]

A simple way to see the turning point in the history of physics is to compare the pace of its development. The most important Galileo's predecessors – Aristotle and Archimedes – lived two thousand years earlier, while Galileo was the most important predecessor to his students and to Newton who was born in the year of Galileo's death. Very important for Galileo was also an astronomical challenge from Copernicus, the acknowledged initiator of "the Scientific Revolution".

But what did Galileo invent which was so entirely new? It was hardly his insights into specific phenomena like inertia and free fall. Insightful geniuses are born rarely but uniformly all over the globe. An Islamic scientist Alhazen (aka Ibn al-Haytham, 965-1040) had an insight into inertia six centuries before Galileo, and a Chinese philosopher Mozi (aka Mo Tzu, 470-391 BC) - twenty centuries before. But those insights were not developed and laid hidden in old manuscripts until historians discovered them.

Joseph Needham (1900-95), a British scientist, historian and famous sinologist, raised the so called Needham Question: "*Why did modern science, the mathematization of hypotheses about Nature, with all its implications for advanced technology, take its meteoric rise only in the West at the time of Galileo? … why modern science had not developed in Chinese civilization (or Indian) but only in Europe?*" This question was sharpen by his realization that "*between the first century B.C. and the fifteenth century A.D., Chinese civilization was much more efficient than occidental in applying human natural knowledge to practical human needs.*" [2]

The key point is not why Europe was the first to launch the modern physics,- somebody has to be the first, - but why for so long nobody joined it beyond Europe. European culture borrowed important innovations from China, India, and Islamic world, like paper, Hindu-Arabic numerals, and algebra. However the greatest Western innovation of the modern physics did not transfer South-East for centuries. The Needham question is not about an exercise in counterfactual history, it is about beneficial cultural infrastructure for science and technology.

The problem of the Scientific Revolution attracted historians since 1930s, with various factors and facets explored and emphasized, but up to now there is no convincing explanation why it happened. [3] The quest was started by Marxist scholars in the wake of two revolutions - the quantum-relativistic one in physics and social one in Russia. Marxists searched for laws of history - including laws of revolution - similar to laws of physics. Boris Hessen's paper "The Social and Economic Roots of Newton's Principia" (1931) initiated externalist approach to science by the idea that the early modern physics arose from a social context to meet practical demands of capitalist economy.[4] In the line of externalism, Robert Merton adopted Max Weber's explanation of the flourishing capitalism by the role of Protestant ideology and argued that the latter was especially beneficial to modern physics with its experimentalism as the key feature. On the other hand, Alexandre Koyre, who coined the very term "the Scientific Revolution", claimed that it was brought about by "mathematization of nature" rather than by the experimental method. And, on the third hand, Edgar Zilsel suggested that the modern physics emerged due to early capitalism that urged contacts between the academically trained scholars and superior craftsmen.

The very diversity of these explanations means the absence of a proper one. Needham in his posthumous publication confirmed that his question was still unanswered.[5] Indeed, the greatest achievement of the Scientific Revolution - celestial mechanics - had no practical value for economy; for its originators - Copernicus, Galilei, Kepler, and Newton - both the empirical and mathematical tools were indispensable; two of the four were Protestants, and two were Catholics; in China, without capitalism, there were fruitful contacts between scholars and superior craftsmen.

To look for a new answer to the Needham question we should start by clarifying the essence of the modern physics invented by Galileo. What distinguishes Galileo's physics from Archimedes' one? And what in Galileo's physics was as modern as in Einstein's?

**Modern physics and fundamental physics**

For Archimedes the mathematician and engineer both mathematics and experiment were tools of research. As to Einstein's notion of modern physics, it was graphically depicted in his letter to M. Solovine (7 May 1952):

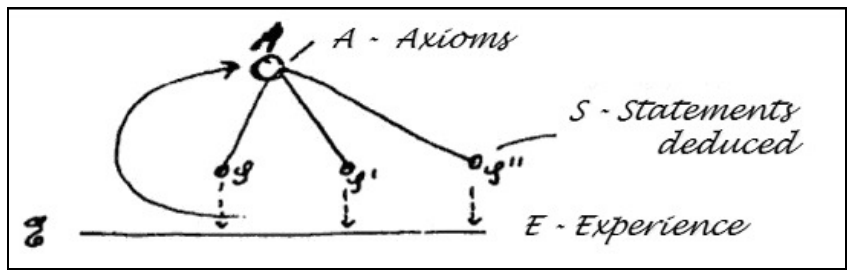

Here fundamental elements of theory (Axioms *A*) are free intellectual inventions arising from empiric Experience *E* by ascending arrow of intuition. Specific exact statements *S* are deduced from *A* to be verified in the *E* (descending arrows).

Here is the difference. All the notions involved in Archimedes' physics are directly evident, while in Einstein's physics the fundamental notions and axioms have not to be evident, - their validations are results of the whole scientific organism. The very first inevident fundamental notion invented by Galileo was "vacuum", or rather "movement in vacuum".

Einstein's scheme can be formulated in the following double postulate of fundamental physics:

1) **There are fundamental Axioms that all the physical laws could be deduced from; those Axioms are not evident, they are as invisible as the foundation stones, or, in Latin, *fundamentum*;**

2) **The human mind is able to probe into this fundamental level of the Universe to understand its working, and any human is free to contribute in the process of this probing and understanding.**

This double postulate defines fundamental science. It was such a fundamental worldview that was the real novelty of the Scientific Revolution starting with Copernicus. As to Galileo, he invented a *specific way to probe into the inevident fundamental level of the Universe's working - by joint use of experiment and mathematical language*. He demonstrated his way of making fundamental science in the notion "movement in vacuum" that resulted in the law of inertia, the principle of relativity, and the law of free fall. Aristotle proved – in more than one way – that there could not be such thing as nothing, aka void, emptiness, vacuum. Galileo never experienced vacuum by his senses, but having based on empirical observation, he employed his brave freedom to invent axioms in mathematical language, his responsibility to experimental verification, and came to a fundamental notion of an "invisible" vacuum that happened to be so fruitful.

Fundamental physics is just a small part of the modern physics – its forefront, the larger part being the old good "Archimedean" physics, where to directly evident notions added are tested and accustomed fundamental ones. However it was the forefront fundamental physics that played the role of powerful engine to propel the rest of physics by providing it with new basic "words" to describe physical reality.

Of course a prerequisite for doing good science is a personal curiosity. Fundamental science requires an extraordinary curiosity, because it is most profitless, with its only gain being new knowledge about the inner workings of the Universe. Galileo was the first fundamental physicist.

## What was the source of the originators' faith in fundamental laws?

What, besides great talents, helped Galileo as well as Copernicus, Kepler and Newton to become originators of the fundamental science? Einstein wrote about Kepler, who *«lived in an age in which the reign of law in nature was by no means an accepted certainty. How great must his faith in a uniform law have been, to have given him the strength to devote ten years of hard and patient work to the empirical investigation of the movement of the planets and the mathematical laws of that movement, entirely on his own, supported by no one and understood by very few!»* [6]

All the originators shared such a faith in fundamental laws, but what was the source of this faith? Most probably, it was the genuine religious faith which all the four greats did have as well. Connection between these two kinds of faith was discovered by a Marxist historian - and, sure, atheist - Edgar Zilsel in his seminal research "The Genesis of the Concept of Physical Law" (1942), where he showed that the expressions "Physical Law" and "Law of Nature" emerged in the 17th century within Biblical worldview by transforming an idea of the Universe governed by laws decreed by God into a basic notion of scientific discourse. [7]

Galileo, who discovered the first fundamental law of modern physics – the law of free fall - never used terms like "Physical Law" and "Law of Nature" in his original Italian publications, instead he used terms like "*ragione*" (proportion, ratio) or "*principio*" (principle). But in his theological letters to his disciple and to his patron (of 1613-15) Galileo started the transformation. [8] Here is a summary of his views:

*«The Scripture and Nature both derive from God, the Scripture as His dictation, the Nature as the obedient executrix of God's commands. The aim of the Scripture is to persuade humans of those propositions which are necessary for their service to God and salvation. To adapt to the understanding of unlearned people, the Scripture speaks many things which differ from the bare meaning of words, and it would be blasphemy to accept them literally by attributing to God hands, human feelings like anger, regret, and forgetfulness.*

> ***Nature**, on the other hand, **never transgresses the laws imposed upon her**, and does not care a whit whether her **abstruse reasons and modes of operation** are understandable to humans. God has furnished us with senses, language, and intellect not to bypass their use and give us by other means the information we can obtain with them. Therefore, whatever sensory experience and necessary demonstrations prove to us concerning natural phenomena it should not be questioned on account of Scripture's words which appear to have a different meaning. This is especially so for the phenomena about which we can read only very few words. The Scripture does not contain even the names of all the planets, and so it was not written to teach us astronomy».*

This worldview in fact contains the double postulate of the fundamental science. There are unbreakable laws governing all the abstruse reasons in Nature, and humans have free abilities to understand those reasons by means of their senses, language, and intellect.

Galileo's "laws imposed by God upon Nature" by the end of the 17th century transforms into just "laws of Nature", and, as Zilsel found, this transformation was made mainly by deeply religious scientists like Descartes and Boyle.

It was the last publication of the 50-yrs old Zilsel in the series published after he arrived to the USA having escaped from Austria occupied by Nazis. In 1944 Zilsel committed suicide, so we don't know what he was thinking about the gap between his earlier Marxist explanation of the Scientific Revolution and his own hint to an anti-Marxist explanation.

The very role of Biblical worldview in the thinking of the first modern physicists is more important than the specific phrase "law of Nature". An atheist Zilsel wrote about "the law-metaphor originated in the Bible", but in a religious worldview most ways to talk about God are metaphorical.

There is no much sense to discuss specific denominations of the four originators. While being genuinely religious they were no less genuinely free and independent in their thinking both in science and in religion, which resulted in their being at odds with theological officialdoms. The only source of religious authority they did not question was the Bible although they felt free to question its interpretations including translations from the original. It is most clearly manifested by Newton's dissertation on corruptions of Scripture. For the four originators the Bible was as original reality as the Nature was.

The connection of the two faiths in the originators' minds made the double postulate of the fundamental science to resonate with the most general postulates the Bible teaches – acknowledgement of the supreme Creator-Lawgiver and the personal freedom of a human endowed by Himself, who has created man in His image and likeness. Regardless of religious diversity in Europe this Biblical postulate had been instilled and dissolved in European culture, which more properly could be named the Biblical culture, since the Bible is the most common element of European sub-cultures, as different as Finnish and Italian, Russian and British. Legitimate offspring of the Biblical culture includes unaffiliated believers as well as atheists, since the freedom of conscience, or separation of state and church were introduced into Western social agenda by profoundly biblical people like the Puritans.

One should not think that in that old time there was no atheists at all. Atheism was well established back in the time of Archimedes, - in Epicureanism. An overt atheist was a colleague and friend of Newton – Edmond Halley, and there were many more covert atheists. But an historical fact is that there was no atheists among the originators of modern physics. As far as their attitude toward the Bible is concerned they would probably agree with the next great fundamental physicist Maxwell, who in the very beginning of his scientific career wrote to his close friend:

*"Now, my great plan<> is to let nothing be wilfully left unexamined. Nothing is to be holy ground consecrated to Stationary Title, whether positive or negative. <> Christianity—that is, the religion of the Bible—is the only scheme or form of belief which disavows any possessions on such a tenure. Here alone all is free. You may fly to the ends of the world and find no God but the Author of Salvation. You may search the Scriptures and not find a text to stop you in your explorations."* [9]

In the same free spirit, a quarter century later, Maxwell declined the invitation to join a society whose stated aim was to defend "the great truths revealed in Holy Scripture ... against the opposition of Science falsely so called." He explained: *"I think that the results which each man arrives at in his attempts to harmonise his science with his Christianity ought not to be regarded as having any significance except to the man himself, and to him only for a time, and should not receive the stamp of a society. For it is of the nature of science, especially of those branches of science which are spreading into unknown regions to be continually —"*.[10] Here the draft ends, but one can guess that Maxwell was to continue something like "… to be continually asking new questions and questioning accustomed answers".

In the next generation the fundamental double postulate was laconically formulated by Einstein: *"Subtle is the Lord, but malicious He is not"*. Having undergone deep religiosity as a child, adult Einstein was far away from any churchlike life, but presenting his "cosmic religion" he wrote that *"the beginning of cosmic religious feeling already appear... in many of the Psalms of David and in some of the Prophets"*.[11]

While in Einstein time one could have a faith in fundamental physical laws just because quite a few laws had been discovered, it was not so in the 16th century when there was none. Hence it seems to be the key fact that all the originators of the modern physics were Biblical theists, whose fundamental worldview was supported by their Biblical "pre-physics" (to avoid words "prejudgment" and "metaphysics").

Supportive parallels worked in fundamental role of non-evident laws in ethics and physics, in Galileo's notion of the two Great Books by the same Author – the book of Scripture and the book of Nature, and in problems of proper translation of these books from the original into vernacular and into the language of everyday life. Hence Galileo's famous words about the book of Nature written in the mathematical language.

## A physicist by the Tree of Knowledge

The hypothesis that the Bible had inspired the Scientific Revolution is a simple answer to the Needham question since the Bible did distinguish European culture from all the others. To show that this answer is not too simple, let's deal with most evident doubts.

By the time of the Scientific Revolution the Bible was around for many centuries. What had it been waiting for so long to launch the modern physics? It waited for the time when the Book took the most prominent cultural role in its history due to Gutenberg and Reformation.

The Scientific Revolution overlapped with a few major European phenomena: Renaissance, Reformation, Capitalism, and political fragmentation of Europe, as impossibility to create All-European Empire or even a predominant European Empire. All these phenomena mattered for the Scientific Revolution, and all of them required sufficient freedom per capita together with responsibility as respect to the rule of law, first and foremost to the rule of the supreme law - intelligible, rational, even if super-rational, but not irrational law. The most fundamental all-European base for such a cultural infrastructure was the Bible whose role started to strengthen since 12th century with the principle "Sola Scripture" (by scripture alone) triumphed in the Protestantism.

It is easy to imagine how the Bible could teach a person endowed with entrepreneurial or artistic talents to be bravely free and responsible, - it's the Parable of the talents (Matthew 25:14-30). But how the Bible could teach an adolescent endowed with a talent to seek knowledge about Nature, but who, like Galileo and Kepler, was willing to become a priest?

Let's look at the Bible through the eyes of not a theologian but an adolescent monotheist endowed with extraordinary curiosity and independent thinking. The very first chapters of the Bible contain the story most related to the situation of fundamental scientific inquiry – the story around the Tree of Knowledge of Good and Evil.

First, our independent smart reader would grasp that this story is about knowledge in general, for to do good in a specific situation one have to know the working of this situation, for example, to know what would heal the specific sickness of a suffering person.

Second, our reader would grasp that God's words not to eat the fruit from the Tree of Knowledge lest to become mortal, was not an outright prohibition but a kind of forewarning, for God's prohibiting commandment ("Thou shalt not …") didn't refer to consequences of its violation.

And, third, a potential researcher of nature could easily comprehend the main motivation of Eve to eat the fruit from the Tree of Knowledge because of longing for knowledge. Since it was the very first action of Eva, apparently both her freedom of choice and her longing for knowledge had been provided to her by her Creator. Hence a potential fundamental scientist would feel free to be dying for a new piece of knowledge about the Universe and would accept his responsibility for this.

Even within Biblical cultural domain there is evidence of such a support from the Bible since its role is somewhat different in different Christian denominations, being most prominent in Protestantism. The idea that Protestant ideology was beneficial to the science was based on the statistical fact of the disproportionately high contribution of Protestants to science since 17th century. [12] This fact is supported by statistics of religious background of Nobel laureates (1901-2002): those with Catholic background are 9% of Nobel laureates and 17% of the world population, while those with Protestant background are 30% of Nobel laureates and 7% of the world population.[13] Such a difference is a result of a few centuries of stronger development of science in the Protestant world, while in the early 17th century science in the Catholic world was no weaker than in the Protestant one. Such a developing disparity is to be explained.

Despite all the differences within Biblical civilization, much greater was the difference of other civilizations in ability to adopt modern science. Instructive is comparison between Russian and Chinese Empires.

China was much more advanced with her "Four great inventions" in technology, traditions in philosophy and astronomical observations. The modern science was brought to China by Jesuit missionaries back in 17[th] century and was welcomed by Chinese emperor, who appointed a missionary his scientific adviser.

More than a century later, the modern science was brought to Russia due to Peter the Great. It started with bringing outstanding scientists from Europe, including Euler and Bernoulli. However very soon a powerful indigenous figure of Lomonosov appeared, and the first world-class results in exact sciences - Lobachevski's geometry and Mendeleev's periodic table of elements – emerged a century before Chinese contribution into the modern science. So strong disparity could be explained by the fact that Russia belonged to Biblical civilization despite all the socio-economical differences with Western Europe. As far as the tiny minority of potential scientists in Russia is concerned, they relied on the same cultural infrastructure as the same minority in the Western Europe.

In China the main hindrance was apparently the lack of a concept of "law of nature". When a missionary tried to explain to Chinese *"that God, who created the universe out of nothing, governs it by general Laws, worthy of his infinite Wisdom, and to which all creatures conform with a wonderful regularity, they say, that these are high-sounding words to which they can affix no idea, and which do not at all enlighten their understanding. As for what we call laws, answer they, we comprehend an Order established by a Legislator, who has the power to enjoin them to creatures capable of executing these laws, and consequently capable of knowing and understanding them. If you say that God has established Laws, to be executed by Beings capable of knowing them, it follows that animals, plants, and in general all bodies which act conformable to these Universal Laws, have a knowledge of them, and consequently that they are endowed with understanding, which is absurd."*[14]

Some Merton's followers pointed out the role of "biblical worldview" in the rise of modern science by focusing on active, optimistic "Christian empiricism" as the new respect to the facts of nature rather than to old-fashioned intellectual speculations. [15] However such empiricism was characteristic only for Protestant ideology and insufficient for the first major triumphs in astronomy and physics authored both by Catholics and Protestants. And this approach had nothing to explain specific intellectual contribution into the new – modern – science.

The approach suggested here starts with selecting astronomy and physics as the core of the Scientific Revolution. The "Biblical worldview" is reduced to two very basic postulates on an invisible super-intelligent Creator-Lawgiver who endowed humans with the free intelligent ability to understand His will and deeds. Those postulates were intellectual and inspirational support for a religious scientist to make fundamental science. The source of those postulates was the Bible itself whose role climaxed in the era of Reformation in both Catholic and Protestant parts of Europe. It did climax regardless of the difference in roles of the Bible in the Catholic tradition and the Protestant innovation, and due to the very fact of debates around the role of the Bible together with its mass printing and translations into vernaculars. As far as science is concerned, the basic theological prototype of the double postulate was much more important than theological subtleties which are too far from scientific problems.

After the debates of Reformation had been over, Protestant innovation became a new tradition, and different roles of the Bible started to work differently in Catholic and Protestant populations of Europe. The Protestant tradition promotes a much more active role of the Bible in daily life that could explain more active development of science than in Catholic and Orthodox traditions.

As to the science beyond physics, it was the very striking success of fundamental physics in demonstrating the ability of the human mind to probe and understand Nature that became a great inspiration and encouragement for scientists the non-physicists.

Since the modern physics was invented in the time when the Bible's social role was the greatest in the European history, no wonder if contribution of the Bible into the genesis of the modern science was no less than the Bible's contributions into European languages and literatures. In Europe even hard-core atheists have to use biblical phrasing. And those Westerners who are open to Eastern wisdoms of Zen, Yoga and so on, implicitly follow the cultural genome of Biblical-European civilization, with its powerful genes of openness, activity, and higher personal freedom and responsibility per capita.

## Why Galileo didn't discover universal gravitation?

Yet the biblical predisposition was not absolutely beneficial. According to a Russian proverb and unregistered law of dialectics there is no bad without some good and no good without some bad.

The very first discovery in fundamental physics, made by Galileo, - the law of free fall - was also the first discovery in physics of gravity. It was the starting point for Newton's law of universal gravitation a few decades later. Was it possible for Galileo himself to discover the law of universal gravitation at his level of mathematization and by his style of doing science?

Yes it was, although Galileo's predisposition was very unfavorable, since he rejected statements on attraction as an explanation of the Solar system. But nevertheless Galileo could come to the law of universal attraction by a way starting with his discovery that free falling object is moving on parabolic trajectory. He understood that parabolic trajectory was but an approximate result for "flat Earth", or for small initial velocity. He didn't know the form of trajectory in general case but he would grasp that very high initial horizontal velocity would make the object to go far away from the Earth.

Galileo is often reproached for his keeping to "backward" ideally circular planetary orbit despite the observational reality summarized in Kepler's law of elliptical orbits. Galileo ignored rather than reject Kepler's laws of planetary motion. Circular planetary orbit was the simplest model to probe into physics of planetary motion, and Galileo could do this. Even without knowing the general form of trajectory of free falling object, he could ask what initial horizontal velocity $V$ would make the object to move at the same distance from the surface of the Earth. And he could answer this question by means of math no more sophisticated than the theorem of Pythagoras: $V = (gR)^{1/2}$, where $g$ is the acceleration of free fall and $R$ is the radius of the Earth.

The motion at a constant distance from of the Earth resembles the Moon's motion too much for Galileo to miss this resemblance. But Galileo would find that the relation $V = (gR)^{1/2}$ holds for the Moon only if acceleration of free fall on the Moon's distance from the Earth $g_M$ is about 400 times less than $g_E$ he had measured on the Earth, while the distance to Moon $R_M$ is about 60 times more than $R_E$. It would hint at relation $g(R) \sim R^{-2}$.

Combining it with the previous, he would get a relation for the astronomically observable parameters $V \sim R^{-1/2}$ .

Having verified this relation for the planets in the Solar system and for the satellites of Jupiter, Galileo would realize that he got the 3rd Kepler's law of planetary motion ***as a result of universal attraction that he had studied in the phenomenon of free fall***.

Thus Galileo would come to the law of universal gravitation. Why didn't he do it ?

A probable reason was his religiosity. Being quite serious about his biblical worldview – regardless of how far he was from official theology – Galileo was unable to accept a quite friendly suggestion from the Pope (who was his admirer) to write about his science freely but without claiming that his theory was real truth rather than a hypothesis, even if the best one compare to others. If Galileo had been an atheist he could condescend to scientific backwardness of the religious authority, and in his writing to address to his colleagues the scientists with obviating repercussions by proper hypothetical wording. But being an honest biblical believer he had to defend his truth-seeking. So he invented a literary form to obviate administrative restrictions in his Dialogues and had to spend too much time and effort for his kind of popular-science writing. Nevertheless he failed to circumvent the scientific ignorance of society and the Church and, as a result of persecution, his intellectual and social freedom was harshly restricted for the rest of his life.

Of course in the history of science Galileo's "unnecessary" popular-science writings played a very important role in propagating the new method of doing science all over Europe. But being too busy for too long with such writings and with opposition to ideological officialdom Galileo left the honor for developing his research on gravity to Newton.

## Scientific progress and intellectual freedom

In thinking about the beneficial cultural infrastructure for scientific progress, comparative history might be a good resource, although in the 21st century to promote the Bible might be not the only way to ensure such an infrastructure. Thanks to the amazing advancement of science, nowadays the double postulate of fundamental science is self-evident without biblical support.

Various social forces are working in a society to find out persons endowed with extraordinary curiosity and independent thinking and to provide them with necessary freedom to develop their personal abilities to make new inventions in science and technology. Here, as the history of science shows, the intellectual freedom is the most relevant among all the human rights. Instructive is the comparison of two sciences – biology and physics – against the same totalitarian background of Stalin's Russia.

In the 1920s and the early 1930s both sciences were doing pretty well in the USSR, until the late 1930s when Stalin's Great Purge killed millions of innocent people including some of the best physicists and biologists. [16] However in the post-war USSR the fates of the two sciences were quite different.

An agronomist T. Lysenko, enthroned in the Soviet biology directly by Stalin, effectively suppressed intellectual freedom in Soviet biology to result in its major destruction.

On the other hand, the urgent need for nuclear weapon made Soviet leaders restrain their control over the intellectual freedom of physicists, and the highest level of the freedom was allowed at the closed nuclear center where nuclear weapons were designed.

One of the weapons designers, the theoretical physicist Andrei Sakharov, in 1968, was expelled from this center after he had written an article "Reflections on Progress, Peaceful Coexistence, and Intellectual Freedom". [17] It transformed the secret "father of the Soviet H-bomb" into a public figure. His way to publicity was unique. Being a top expert in strategic balance and privy to strategic information in full, in 1967, he became gravely concerned with a problem of strategic antiballistic defense. He sent a secret detailed letter to the Politburo explaining the increased threat of nuclear war. In those days he felt himself a defender of socialism and a non-dogmatic Marxist. Sakharov saw the fact that the founders of Marxism didn't foreseen: due to advancement in science and technology humanity was facing the threat of global suicide within half an hour, the travel time for a nuclear missile. Sakharov actions were exercises in intellectual freedom coupled with social responsibility. He was well aware that the real intellectual freedom could thrive only on the basis of respect for the rule of law.

However, the Soviet leaders had no respect for both intellectual freedom and social responsibility of citizens, they did not heed the advice of a top non-dogmatic expert and didn't allow Sakharov to publish a non-secret version of his analysis. It was only then he found that his intellectual freedom, so essential in his profession, was dangerously restricted. Feeling himself free enough, he went public to prevent nuclear war. Correcting the official formula, Sakharov wrote that "*evolution, not revolution, is the best locomotive of history*" and confessed himself to be a "*reformer and principled foe of violent revolutionary changes of the social structure, which have always led to the destruction of the economic and legal system, to mass suffering, lawlessness, and horror.*"

Soviet leaders failed to make the necessary reforms, and the regime collapsed.

Chinese economic reforms show that it was not the only possible outcome. And if Chinese reformers could also create a beneficial cultural infrastructure for scientific inventiveness it would be the best practical response to the grand question of Needham, who was so sympathetic to Chinese civilization. History hints that respect for intellectual freedom and for the rule of law is the best secular approximation to the Biblical prerequisites for fundamental physics at the time of its origin.

### *Acknowledgment*

I am grateful to Lanfranco Belloni, Robert S. Cohen, Diederick Raven, Chia-Hsiung Tze, Sergey Zelensky, and the Methodological seminar at the Institute for History of Science and Technology (Moscow) for stimulating discussions.

---

[1] A. Einstein. On the Method of Theoretical Physics, 1933.

[2] J. Needham, The Grand Titration: Science and Society in East and West, Toronto: University of Toronto Press, 1969, pp. 16, 190.

[3] H. F. Cohen, The scientific revolution: a historiographical inquiry. Chicago: University of Chicago Press, 1994.

[4] Gessen (Hessen) B. M. Socialno-ekonomicheskie korni mekhaniki N'yutona. M.-L., GTTI, 1933. 77 pp. English translation in: Gideon Freudenthal and Peter McLaughlin, The Social and Economic Roots of the Scientific Revolution, Springer, 2009, pp. 41-101.

[5] Joseph Needham. Foreword. In: Edgar Zilsel. The Social Origins of Modern Science. Ed. Diederick Raven, Wolfgang Krohn, and Robert S. Cohen. Dordrecht: Kluwer Academic Publishers, 2000.

[6] A. Einstein. Johannes Kepler (in commemoration of the 300th anniversary of his death), 1930.

[7] Edgar Zilsel. "The Genesis of the Concept of Physical Law" // The Philosophical Review, May 1942, p. 245-279 <http://www.compilerpress.ca/Competitiveness/Anno/Anno%20Zilsel%20Genesis.htm>; In: Edgar Zilsel. The Social Origins of Modern Science. 2000.

[8] [Galileo Galilei. Letter to Benedetto Castelli (1613)](http://www.disf.org/en/documentation/03-Galileo_PBCastelli.asp) <http://www.disf.org/en/documentation/03-Galileo_PBCastelli.asp>

[Galileo Galilei. Letter to Madame Christina of Lorraine, Grand Duchess of Tuscany Concerning the Use of Biblical Quotations in Matters of Science (1615)](http://www.disf.org/en/documentation/03-Galileo_Cristina.asp) <http://www.disf.org/en/documentation/03-Galileo_Cristina.asp>

[9] L. Campbell and W. Garnett, *The Life of James Clerk Maxwell*, 2nd ed. London: Macmillan, 1884 <http://www.sonnetsoftware.com/bio/maxbio.pdf>, p. 96.

[10] Ibid, , p. 196.

[11] Quote in M. Jammer. Einstein and Religion: Physics and Theology. Princeton University Press, 2011, p. 76, 234.

[12] H. F. Cohen, The scientific revolution: a historiographical inquiry. Chicago: University of Chicago Press, 1994, p. 314.

[13] J. M. Rector and K. N. Rector. What is the Challenge for LDS Scholars and Artists? // Dialogue - A Journal of Mormon Thought, 2003, Vol. 36, N 2, p. 34-46.

[14] Quote in: H. F. Cohen, The scientific revolution, 1994, p. 467.

[15] H. F. Cohen, The scientific revolution, 1994, ch.5.

[16] G. Gorelik, and V.Ya.Frenkel, Matvei Petrovich Bronstein and Soviet Theoretical Physics in the Thirties, Basel-Boston: Birkhaeuser Verlag, 1994; Springer Basel AG, 2011 (e-book)).

G. Gorelik, 'Meine antisowjetische Taetigkeit...' Russische Physiker unter Stalin, Transl. H. Rotter. Braunschweig/Wiesbaden: Vieweg, 1995.

[17] G. Gorelik, with A. W. Bouis, The World of Andrei Sakharov. Oxford University Press. 2005.